\documentclass[12pt,a4paper]{article}

\usepackage{graphicx}
\usepackage{subfig}
\usepackage{textcomp} 

\begin{document}

\title{\textbf{HERBIG-HARO FLOWS AND YOUNG STARS IN THE DOBASHI 5006 DARK CLOUD}}
\author{T. A. Movsessian$^1$, T. Yu. Magakian $^1$, A. S. Rastorguev$^2$ \\
and H. R. Andreasyan$^1$}

\date{\today}
\maketitle

\begin{abstract}
    Two new Herbig-Haro flows were found in a study of the isolated Dobashi 5006 dark cloud (1$= 216^\circ.7$, b= $-$13$^\circ.9$): one certain (HH 1179) and one presumable, associated with the infrared sources 2MASS 06082284$-$0936139 and 2MASS 06081525$-$0933490, correspondingly. Judging from their spectral energy distributions, these sources may be Class 1 objects with luminosities of order 23 $L_{\odot}$ and 3.6 $L_{\odot}$ , respectively. They are part of the poor star cluster MWSC 0739, study of which based on data from the Gaia DR3 survey has made it possible to detect 17 stars which are probably members of it. A list of them and their main parameters is given. The distance of the cluster is estimated to be 820 pc and the color excess on the path to the cluster is E(BP-RP)$\approx$1.05 mag. All of these stars are PMS-objects and most of them are optically variable. It is concluded that the newly discovered compact star-formation zone in the Dobashi 5006 cloud has an age of no more than a few million years and this process continues up to the present time.

\end{abstract}


\textbf{Will be printed in Astrophysics, 2023, No.1}

$^1$ V. A. Ambartsumyan Byurakan Astrophysical Observatory. National Academy of Sciences of the Republic of Armenia; e-mail: tigmag@sci.am; tigmov@web.am

$^2$ Physics Faculty, Lomonosov Moscow State University. Moscow, Russia

\section{Introduction}
One of the basic signs of an active star-formation process in dark clouds is the presence of Herbig-Haro objects. On the other hand, the presence of
 infrared (IR) sources with a continual spectrum typical for young stellar
objects, deeply embedded in a cloud, as well as the presence of stars associated with compact  reflection  nebulae, may be regarded as similar indicators..

Searches for and studies of Herbig-Haro objects have already been conducted at the Byurakan Observatory for several decades. A new narrow-band survey of dark clouds with the 1-m Schmidt telescope and narrow-band light filters, the Byurakan Narrow Band Imaging Survey (BNBIS), has been initiated in recent years. Initially, searches were mainly carried out in those dark clouds that contained stars associated with so-called cometary reflecting nebulae. Later the list of targets for the survey was supplemented by dark clouds with embedded bright infrared sources. The first results of this survey, relating to the star-formation regions Mon RI and LDN 1652, were previously published in Refs. 1 and 2.

In this article we present the results of a study of objects inside the Dobashi 5006 dark cloud [3]. Together with the neighboring Dobashi 5007, this dark cloud forms a small isolated group with galactic coordinates 1$= 216^\circ.7$, b= $-$13$^\circ.9.$

Inside the Dobashi 5006 cloud there is a group of nebulous stars which was included as MWSC 0739 [4] (see Fig. 1) in the catalog of star clusters in the Milky Way. Probably the first attention to this field was drawn in the paper by Gyulbudagyan et al [5], where the nebulous object HHL 35a was noted. Because that article does not include a description and identification chart, it is impossible to indicate precisely which of the nebulous objects in which this region is so rich corresponds to the presumed Herbig-Haro object designated as HHL 35a by the authors of Ref. 5. In charts of the WISE survey bright IR sources are observed in this region, and this may also be an indication of
an active star-formation process.       All of this group was of sufficient interest to be included in our observational
program.

\begin{figure}[h!]
  \centering
  \includegraphics[width=8cm]{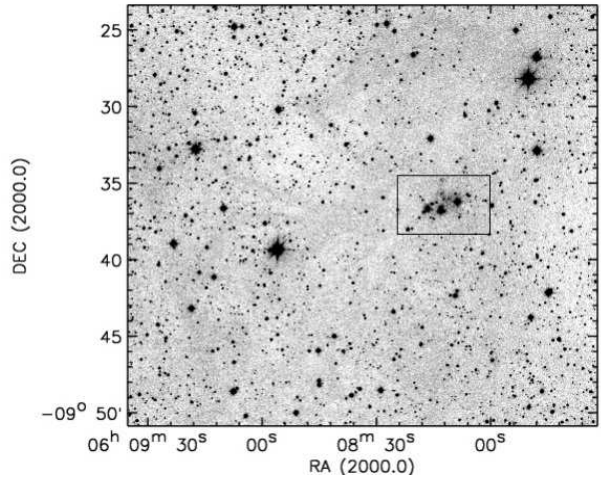}
   \caption{An image of the region of Dobashi 5006
obtained with the Schmidt telescope in the H$\alpha$    line.
The rectangle indicates the cluster MWSC 0739.
}
   \label{Fig1}
\end{figure}

\section{Observations}
The observations were performed on the 1-m Schmidt telescope at the Byurakan Observatory in December 2020 and February 2021. Since the telescope has been completely modernized, it was, in particular, equipped with a Apogee Alta 16M CCD detector, which provides a field of view of about one degree with a pixel size of 0.8683$^{\prime\prime}$ [6].

Narrow-band light filters with a 100 A passband and central wavelengths of 6500 and 6760 \AA\  were used in the observations to obtain images in the H$\alpha$ and an [SII] line, respectively. A medium band filter at 7500 \AA\ with a passband of 250 \AA\ was used to obtain images in the continuum.

During the observations in each filter five-minute exposures were made with shifting of the images to smooth out variations in the sensitivity of individual pixels. The combined effective exposure in the H$\alpha$ filter was 5400 s, in the [SII] filter, 8400 s, and in the continuum, 1800 s. The images were processed by the standard method using a specially developed IDL program package that includes subtraction of bias and dark image, removal of traces of cosmic particles, and correction of vignetting.

Searches for Herbig-Haro objects as such were carried out in the classical method developed by van den Bergh [7] by comparing images in the H$\alpha$ and [SII] emissions with images obtained in the continuum spectrum. Experience shows that in the overwhelming majority of cases this method is sufficiently effective for identifying Herbig-Haro objects.

\section{Results}

\subsection{Herbig-Haro flows}
A study of the images we obtained revealed several emission objects in the region of the group MWSC 0739. Thus, to the south of the center a group of three objects were detected which are clearly visible in the H$\alpha$ and [SII] lines, but are absent in the image obtained in the continuum (Fig. 2). These knots are undoubtedly Herbig-Haro objects. The ratio of the intensities of the emission lines in the knots is different, which indicates different levels of excitation. The faintest on the whole, knot A, is brightest in the H$\alpha$ line, while
B is brighter in [SII], and in knot C their ratio equals unity. This group is weakly noticeable in the images in the DSS-2 atlas; probably it was the one designated in Ref. 5 as HHL 35a.
An analysis of the images of the region in the 2MASS and WISE surveys shows that these knots are positioned on a straight line with the nearby 2MASS 06082284$-$0936139 source, which is entirely invisible in the optical range. It is extremely probable that it is the source of the given Herbig-Haro flow.

\begin{figure}[h!]
  \centering
  \includegraphics[width=0.8\textwidth]{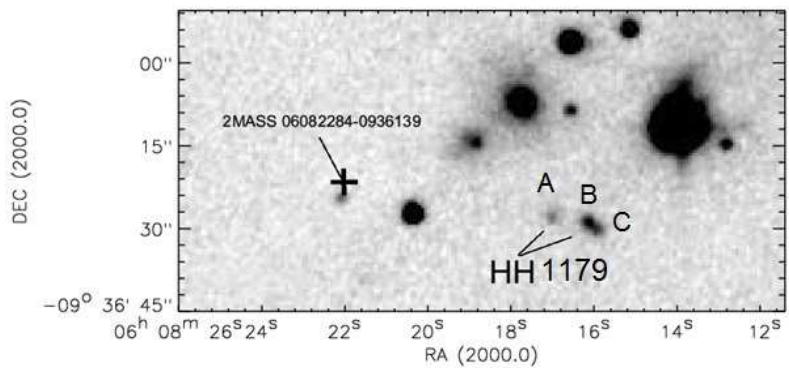}
   \caption{An image of the flow HH 1179 obtained with the Schmidt telescope in the H$\alpha$ line. The individual knots are indicated by letters. The cross shows the position of the IR source 2MASS 06082284$-$0936139.
}
   \label{Fig2}
\end{figure}

To the north of the cluster MWSC 0739 at a distance of 1.5 ang. min from the center there is a reflection nebula
of cometary shape, which is very clearly visible in the z band of the PanSTARRS survey. Immediately inside this
nebula, on its axis, an emission knot was found which is visible only in the H$\alpha$ line (Fig. 3).  Probably it is also
a Herbig-Haro object, but for a final conclusion it will be necessary to obtain an image of this region in the [SII] line in with a deep limit.      Given the characteristic relationship between cometary nebulae and Herbig-Haro objects,
it may be assumed that is should be excited by a central source, which also illuminates the reflection nebula. A comparison of our images with data from the PanSTARRS, 2MASS, and WISE surveys shows that the exciting star (2MASS 06081525$-$0933490) cannot be seen in the visible and is shifted by 1-2$^{\prime\prime}$ to the west from the bright peak of the reflection nebula. This effect, caused by absorption in the circumstellar dust disk, is often observed in similar cases (e.g., HH 83).

\begin{figure}[h!]
  \centering
  \includegraphics[width=0.6\textwidth]{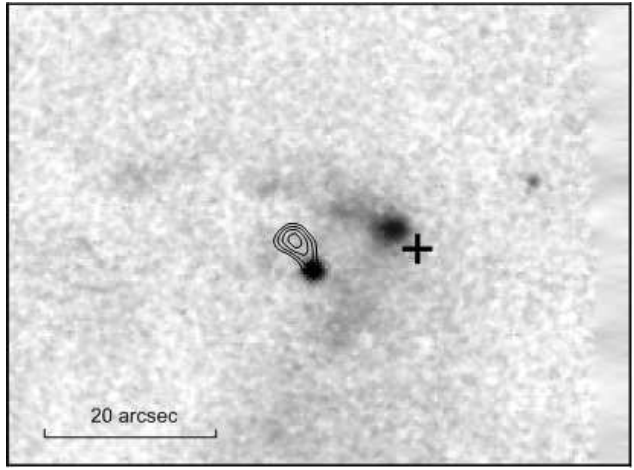}
   \caption{A cometary nebula and the presumed Herbig-Haro object. The half-tones show the image in the continuum taken from the PanSTARRS survey (z band), with superposition of the H$\alpha$ -image obtained after our subtraction of the continuum spectrum (contours). The cross indicates the
position of the source 2MASS 06081525$-$0933490.
}
   \label{Fig3}
\end{figure}

The coordinates of the knots, their visible distances, and position angles with respect to the direction to the assumed sources are listed in Table 1. Also shown there are the distances from the sources calculated for a distance of 820 pc to the object. (The justification for this estimate is given below.)

\begin{table}
\caption{Coordinates of the Knots and their Parameters}
\label{HHcoords}
\centering
\setlength{\tabcolsep}{2pt}
\begin{tabular}{l c c c l l l}
\hline \hline
Object & RA (2000) & Dec (J2000) & Source 2MASS & $r^{\prime\prime}$ & P.A.${^\circ}$ & Dist (pc)\\
\hline
Knot in nebula  &    $06^h 06^m 16.15^s$ &   $-09^\circ 33^{\prime} 47.8^{\prime\prime}$  & \begin{footnotesize} 06081525-0933490\end{footnotesize} &       14  & 87  &    0.06 \\                                        
HH 1179 A   &    06 08 18.43  &  $-$09 36 21.8 &   \begin{footnotesize} 06082284-0936139\end{footnotesize} &   66 & 263 &  0.27 \\
HH 1179 B   &    06 08 17.65  &  $-$09 36 22.23 & -\textquotesingle\textquotesingle-  &     78   &  -\textquotesingle\textquotesingle- &    0.32 \\
HH 1179 C    &   06 08 17.44 &   $-$09 36 24.63 & -\textquotesingle\textquotesingle-  &  82     & -\textquotesingle\textquotesingle-   &     0.34 \\
\hline
\end{tabular}
\end{table}

\subsection{Sources}

For further study of the characteristics of the IR-sources which presumably excite the above described Herbig-Haro flows, we have analyzed their spectral energy distributions (SED).

The source 2MASS 06082284$-$0936139 coincides with WISE J060822.89-093614.2. In addition, at 8$^{\prime\prime}$ to the north from it there is a very noticeable far-IR source AKARI/FIS 0608229$-$093607. Given the angular resolution and accuracy of the coordinates in the mid and far IR range, it is fully obvious that this is one and the same object. Thus, to construct the SED we have use the 2MASS, WISE, and AKARI surveys. Note that this object seems not to have been registered in the IRAS catalogs: the closest to it by coordinates of the IRAS 06059$-$0935 is at a distance of 34$^{\prime\prime}$. An analysis with higher angular resolution in Ref. 8 shows that, in fact, IRAS 06059$-$0935 nevertheless coincides with 2MASS 06082284$-$0936139, but breaks up into a group of three or more objects. Thus, we decided not to use data from the IRAS survey for the SED.

The source 2MASS 06081525$-$0933490, as noted above, cannot be seen in the visible range and becomes clearly noticeable only in the K band of the 2MASS survey. In the mid IR band of the WISE survey, to the west and east from it another two stars are observed that are comparable to it in brightness. An analysis of the coordinates shows that 2MASS 06081525$-$0933490 was recorded in the AKARI/FIS survey as the source 0608149$-$093356. Given the imprecision in the coordinates of this star, it may also be identified with the source IRAS F06058$-$0933. However, on constructing the SED it turned out that the data from the IRAS survey agree poorly with the others, apparently because of the influence of other stars in the field on the measurements. Thus, in this case we also avoided using the IRAS data, and constructed the SED using data from the 2MASS, WISE, and AKARI surveys. The final energy distributions are shown in Fig. 4.

\begin{figure}[h!]
  \centering
  \includegraphics[width=0.7\textwidth]{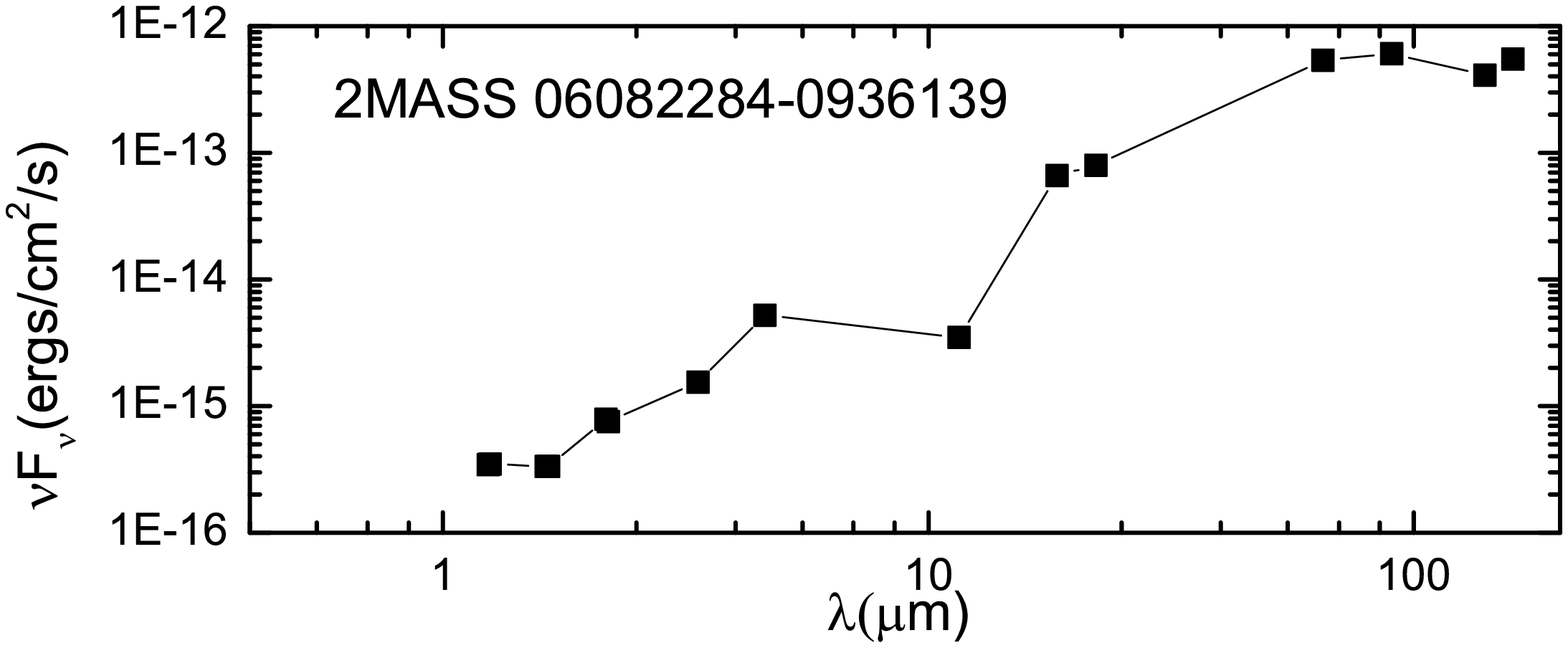}
  \includegraphics[width=0.7\textwidth]{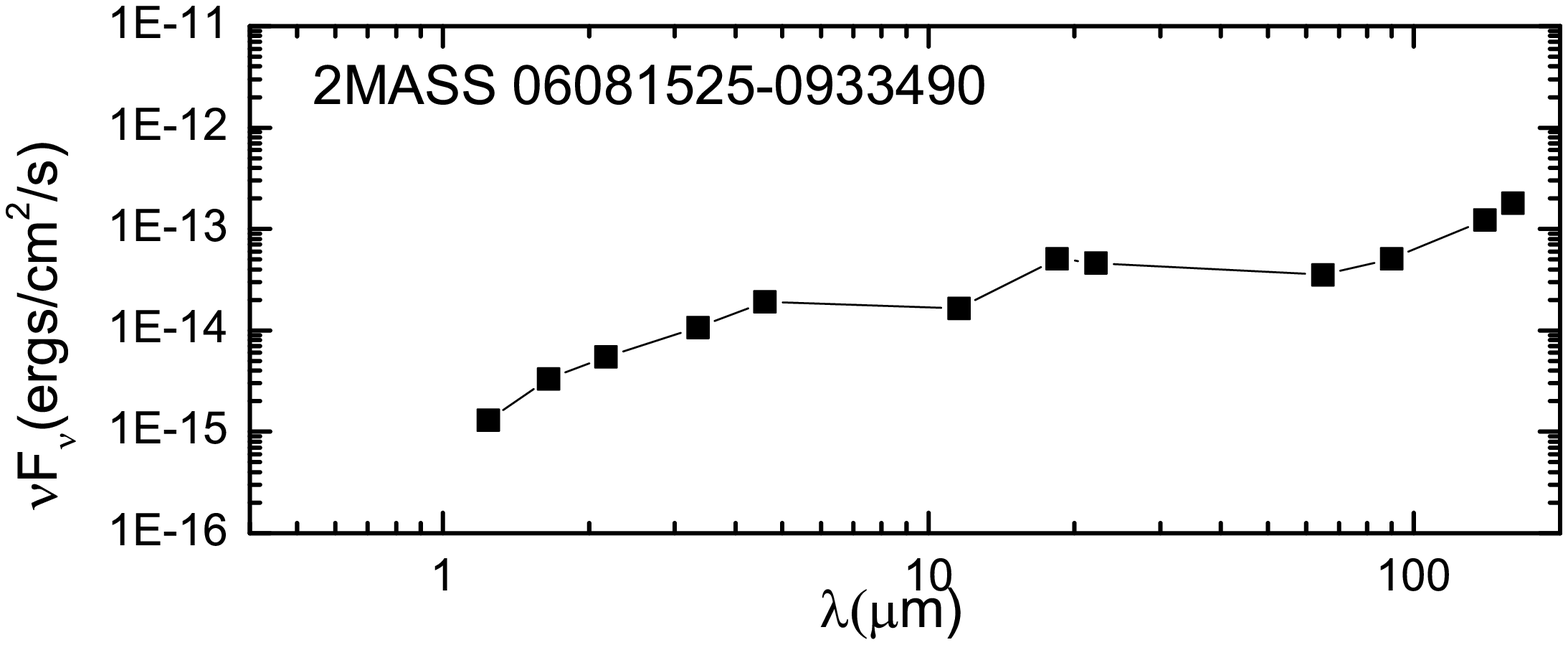}
   \caption{Spectral distributions of the energy of IR-sources presumably associated with Herbig-Haro objects.
}
   \label{Fig4}
\end{figure}

As can be seen from Fig. 4, the SED for the two sources increases monotonically from the near to the far infrared, and in the visible the sources are not observed at all. This all suggests that they belong to Class I IR-sources and are quite young. This is indicated by their noticeable bolometric luminosity, which we calculated for a distance of 820 pc (see below regarding this estimate). For 2MASS 06081525$-$0933490 its lower limit (neglecting interstellar absorption) is 3.6 $L_{\odot}$ , while for 2MASS 06082284$-$0936139, it is 23 $L_{\odot}$.

\subsection{The cluster MWSC 0739.} 

This group of stars was first noted as a possible poor open star cluster in Ref. 9 under the number FSR 1115. It was then included in the general catalog of star clusters in the Milky Way (MWSC) ([4], see Ref. 10, as well) as the object MWSC 0739. Based on the estimates given in these catalogs (which are mainly obtained by pipeline processing of data from the PPMXL and 2MASS catalogs), it is at a distance of 1068 pc, has a radius of the central part of 5$^{\prime}$.4, and of 9$^{\prime}$ for the whole, and must contain 60-70 stars. Its proper motion was also estimated (pmRA = 0.55 ms/yr, pmDE = 3.41 ms/yr) with a color excess of E$_{B-V}$ = 2.353. Since that time, this cluster has not been subjected to detailed study.

Our attention was drawn to the unusual morphology of MWSC 0739, the bulk of which is a chain of ten or more stars mainly with roughly the same brightness which are also clearly visible in the near and mid IR (on the 2MASS and WISE charts). More than 5 stars in this chain are associated with small reflection nebulae and illuminate parts of the dark cloud Dobashi 5006 in which the cluster is embedded. These nebulae are noticeable as well in the near IR in images from the 2MASS survey. We decided to make a more detailed study of the cluster MWSC 0739 using data from the latest survey Gaia DR3.

Measurements of all the stars within a region of radius 6$^{\prime}$ around the center of MWSC 0739 were extracted
from the archive of observational data of which there was a total of 336 objects. The sample of possible members of the cluster was derived, initially for a RUWE criterion (Renormalized Unit Weight Error) of $< 1.4$, and also based on the requirement of closeness of the heliocentric distances and the components of the proper motions of the stars. For example, a similar study (Ref.11), based on data  from GAIA DR2, was made of a young embedded open cluster vdB 130 in the stellar association Cyg OB1 and demonstrated the effectiveness of such approach.

The criteria of closeness of the distances and proper motions are satisfied by 17 stars from the sample. They are shown in the ``parallax-stellar magnitude RP'' diagram (Fig. 5), where the concentration of stars at distances ranging from 800 to 850 pc shows up quite distinctly, as well as in the diagram of their proper motions (pmRA-pmDE) (Fig. 6). A significant concentration of stars with similar proper motions (i.e., tangential velocities) within a limited region of space is a serious argument in favor of the reality of a gravitationally-coupled group of stars (cluster). Data on the radial velocities of these stars are lacking. The average values of the components of the proper motion of the cluster MWSC 0739 are pmRA = $-2.50\pm0.06$ mas/yr and pmDE = $ +0.79\pm0.07$ mas/yr, and the dispersions of the proper motions are, respectively, 0.25 and 0.29 mas/yr for the two coordinates, which for a distance on the order of 800 pc corresponds to a dispersion in the tangential velocity of about 1.5 km/s. Given all the uncertainties, which inevitably increase the estimated dispersion in the velocities, this value corresponds to that which may be expected in young stellar clusters.

\begin{figure}[h!]
  \centering
  \includegraphics[width=0.7\textwidth]{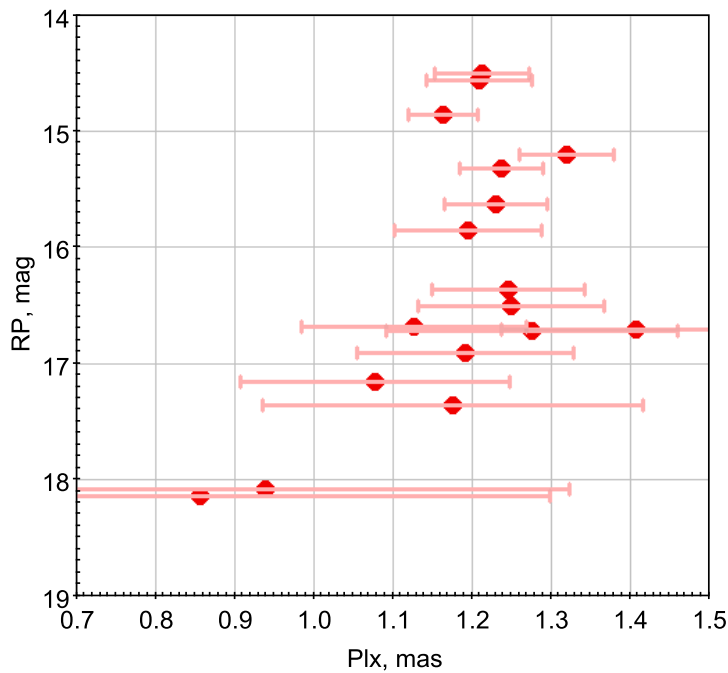}
   \caption{A ``parallax (mas)-visible stellar magnitude RP'' diagram for the stars selected as possible members of the cluster MWSC 0739.
}
   \label{Fig5}
\end{figure}

\begin{figure}[h!]
  \centering
  \includegraphics[width=0.7\textwidth]{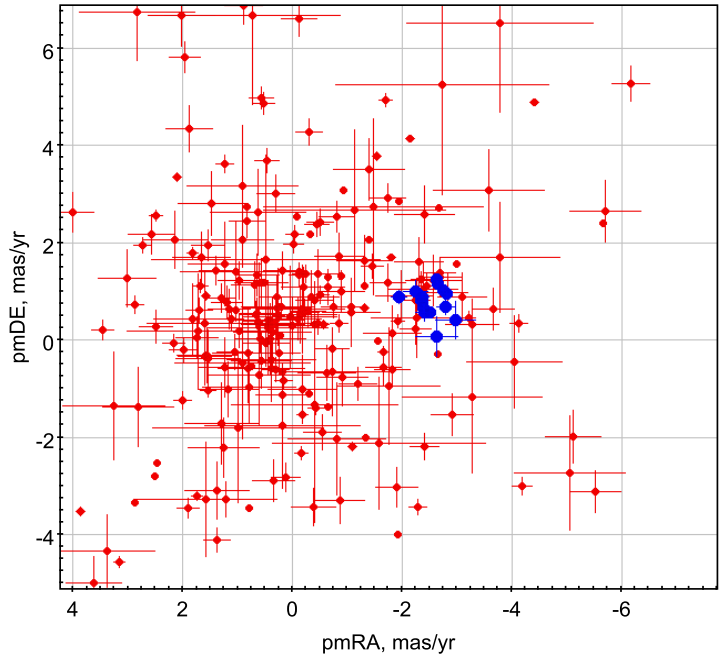}
   \caption{A two-dimensional (pmRA-pmDE) diagram of the proper motions for all the stars in the region MWSC 0739. The compact grouping of the stars with distances of about 800-850 pc, which are the probable members of the cluster (large blue circles) is clearly noticeable.
}
   \label{Fig6}
\end{figure}

Table 2 is a list of the 17 stars which we regard as probable members of the cluster MWSC 0739, with their astrometric and photometric parameters taken from the GAIA (DR3) catalog. Variable, binary, and nebulous stars are pointed out. Figure 7 is a chart of the region of the cluster with the selected stars. These stars include three optically binary stars (see Table 2, in column Bin) with distances of the components of 2$^{\prime\prime}$. The star 3005250640635930880 has RUWE=1.83, but based on other parameters, including traces of a reflection nebula in the surroundings (which may also explain the higher value of RUWE), clearly also belongs to the cluster.

\begin{table}
\caption{List of Stars that are Probable Members of the Cluster MWSC 0379 and their Parameters}
\label{stars}
\centering
\begin{tiny}
\setlength{\tabcolsep}{1pt}
\begin{tabular}{@{} l c c c c c c c c c c c c c}
\hline \hline
N &Gaia DR3 30052... & Plx (mas) & pmRA (mas/yr) & pmDE (mas/yr) & RUWE & G(mag)& BP-RP(mag) & Var & Bin & Neb & [Fe/H] \\
\hline
1  &  49605548347776 &    1.2289$\pm$0.1174 &  $-2.372\pm$0.114 &   0.879$\pm$0.129 &     0.999 &   17.85 &  3.15 &    +  & & &  0.13 \\                    
2  &  49644203766272 & 1.2554$\pm$0.1839 &  $-2.254\pm$0.223 &  0.991$\pm$0.217 &    0.963 &  18.61 &  3.31  &  &   +  &   + & \\      
3 &   49708626938112 & 1.0562$\pm$0.1696 &  $-2.385\pm$0.178 &   0.904$\pm$0.186 &    1.092 &  18.44 &  3.17  &  + & & &                      $-$0.32 \\
4 &   49712921384320 & 1.1882$\pm$0.0662 &  $-2.394\pm$0.072 &   0.645$\pm$0.074 &    0.976 &  16.21 &  2.86  & &          +    &   +  &\\     
5 &   49712922990336 & 1.1424$\pm$0.0444 &  $-2.661\pm$0.048 &   1.153$\pm$0.050  &    1.088 & 16.05 &  2.48 &   + & &              +  &     $-$2.07 \\
6 &   49747282725376 & 1.1704$\pm$0.1370 &  $-1.935\pm$0.144 &   0.876$\pm$0.141   &  0.981 &  18.25 &  3.19 &   +  & & &                     0.48 \\
7 &   49781642475264 & 1.2099$\pm$0.0641 &  $-2.517\pm$0.067 &   0.554$\pm$0.071   &  1.077 &  16.84 &  2.36  &  +  &     +   & & &\\            
8 &   49811707389952 & 0.8363$\pm$0.4418 &  $-2.972\pm$0.372 &   0.397$\pm$0.394   &  1.034 &  19.65  & 3.36 \\                            
9 &   49914785387392 & 1.3869$\pm$0.1708 &  $-2.416\pm$0.172 &   0.562$\pm$0.173  &    1.041 &  18.23 &  4.13  & &       &         +     \\  
10 &  49914785398784 & 1.1068$\pm$0.1421 &  $-2.794\pm$0.153 &   0.679$\pm$0.153  &    0.950 &  18.18 &  4.02  &  &   &         +       \\
11 &  49919081421312 & 1.2996$\pm$0.0591 &  $-2.368\pm$0.062  &  0.789$\pm$0.066  &   1.135 &  16.49 &  3.05 &   +  &   &          +   \\    
12 &  49919081421696 & 1.1748$\pm$0.0937  & $-2.819\pm$0.095  &  0.958$\pm$0.105  &   1.081 &  17.25 &  3.46 &   +  &   &   +       \\
13 &  50606276197376 & 1.2175$\pm$0.0519 &  $-2.756\pm$0.052 &   1.021$\pm$0.058   &  1.074 &  16.47 &  2.49 &   + & & &      $-$0.78\\
14 &  50640635930880 & 1.1923$\pm$0.0594 & $-2.215\pm$0.061 &   0.552$\pm$0.066  &   1.830 &  15.58  & 2.28   & & & &            $-$1.23\\
15 &  50842498376064 & 1.1549$\pm$0.2409 &  $-2.451\pm$0.221 &   1.109$\pm$0.245   &  1.021 &  18.83 &  2.97  & & & &            0.46\\
16 &  74314495656832 & 1.2253$\pm$0.0962 &  $-2.628\pm$0.096 &  1.245$\pm$0.103   &  0.932 &  17.64 &  3.00 &   +  & & &                $-$0.26 \\
17 &  74623733931520 & 0.9177$\pm$0.3838 &  $-2.625\pm$0.381 &   0.0724$\pm$0.407 &  0.991 &  19.50 &  3.17 &    \\

\hline
\end{tabular}
\end{tiny}
\end{table}

\begin{figure}[h!]
  \centering
  \includegraphics[width=0.7\textwidth]{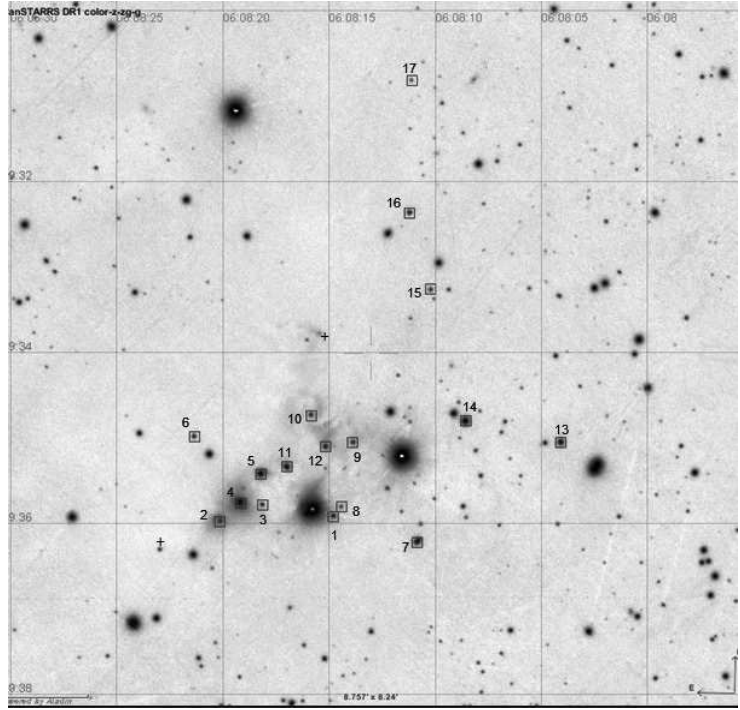}
   \caption{An identification chart for stars in the group MWSC 0739 (denoted by squares) in an image from the PanSTARRS survey. The numbers correspond to positions in Table 2. The ``+'' signs indicate the positions of the IR sources 2MASS 06082284$-$0936139 and 2MASS 06081525$-$0933490.
}
   \label{Fig7}
\end{figure}

After exclusion of two faint stars ($>$18 mag) with large errors in the parallax, the average parallax of the remaining 15 stars equals 1.222$\pm$0.020 mas. Essentially the same value (1.222$\pm$0.09 mas) can be obtained if only the 7 stars which illuminate the small reflection nebulae, i.e., objects that are \textit{a priori} associated with the dark cloud Dobashi 5006, are used for the average parallax. The average distance corresponding to this parallax can be estimated as $<$D$>$ $\approx 818 \pm 14$ pc. The average photogeometric distance calculated for the same objects from the catalog of Ref. 12 is $<$Rpgeo$>$ $\approx 807\pm 21$ pc. However, we believe that when the discussion concerns the distances to stellar clusters, it is methodologically better to make direct use of the catalog values of the trigonometric parallaxes of the stars, since
 Bayesian estimates of the geometric and photo-geometric (Rgeo and Rpgeo [12]) distances are based on \textit{a priori} information on the exponential diminishing of the stellar concentration along the line of sight, which is convenient for the field stars, but does not correspond at all to the distribution of the radial concentration in the cluster, which is better described by a normal law. Thus, we took the distance to MWSC 0739 to be 820 pc, as estimated from the trigonometric parallaxes. Nevertheless, for some estimates in this work we have also used the Rpgeo distances.

For analysis of interstellar absorption in the region of MWSC 0739, data from the latest StarHorse-2 catalog [13], based on a set of multi-color photometric observations from PanSTARRS-1, SkyMapper, 2MASS, and AllWISE, along with data from the GAIA eDR3 catalog, have been used. The estimates of the absorption A$_V$ given there were converted to A$_{RP}$ corresponding to the GAIA photometric system, based on the relationships of the color excesses taken from Ref. 14. Judging from the Rpgeo-A$_{RP}$ diagram, the absorption along the path to the cluster (D $\approx$ 800 pc) is roughly A$_{RP} \approx$ 1.5 mag, while the corresponding color excess is E(BP$-$RP) $\approx$ 1.05 mag. As we can see, all these values (the distance, proper motion, and overall color excess for the cluster) differ very greatly from the estimates in the MWSC catalog. However, in any case it should be noted that inside the cluster itself one can observe a significant differential absorption, mainly created by the dust shells surrounding young stars.

A ``normal color $(BP-RP)_0$ - absolute magnitude M$_{RP}$'' diagram for a distance of 820 pc and an overall color excess E(BP - RP)$\approx$ 1.05 mag (see above) is shown in Fig 8a. It is clear that, as opposed to the chaotic distribution of stars in the field (points), which obviously belong to the distant background, the 17 selected stars of the cluster (circles) are distributed above the main sequence (the isochrones correspond to a solar chemical composition and an age of $log\ t$ = 6.8 and 7.4), which yet again confirms their PMS nature. If, on the other hand we account for the possible additional reddening in the circumstellar shells, using the individual estimates of A$_{RP}$ from the GAIA DR3 data base, then the 8 members of the cluster for which these data are available manifest a substantially lower spread in the color index and a stronger concentration toward the isochrone $log\ t$ = 6.8 (see Fig. 8b).

\begin{figure}
    \centering
    \subfloat[\centering]{{\includegraphics[width=6cm]{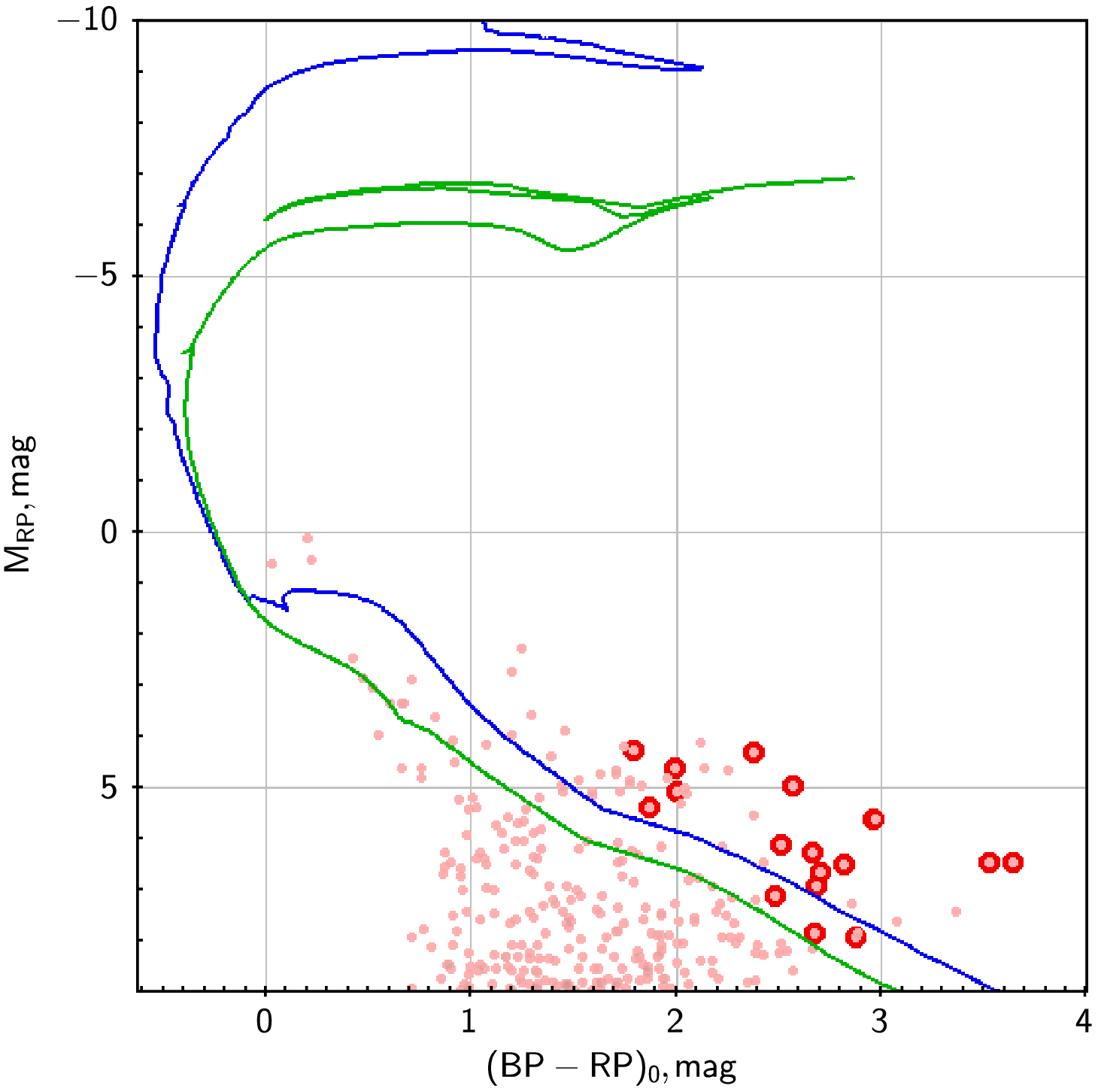}} }
    \subfloat[\centering]{{\includegraphics[width=6cm]{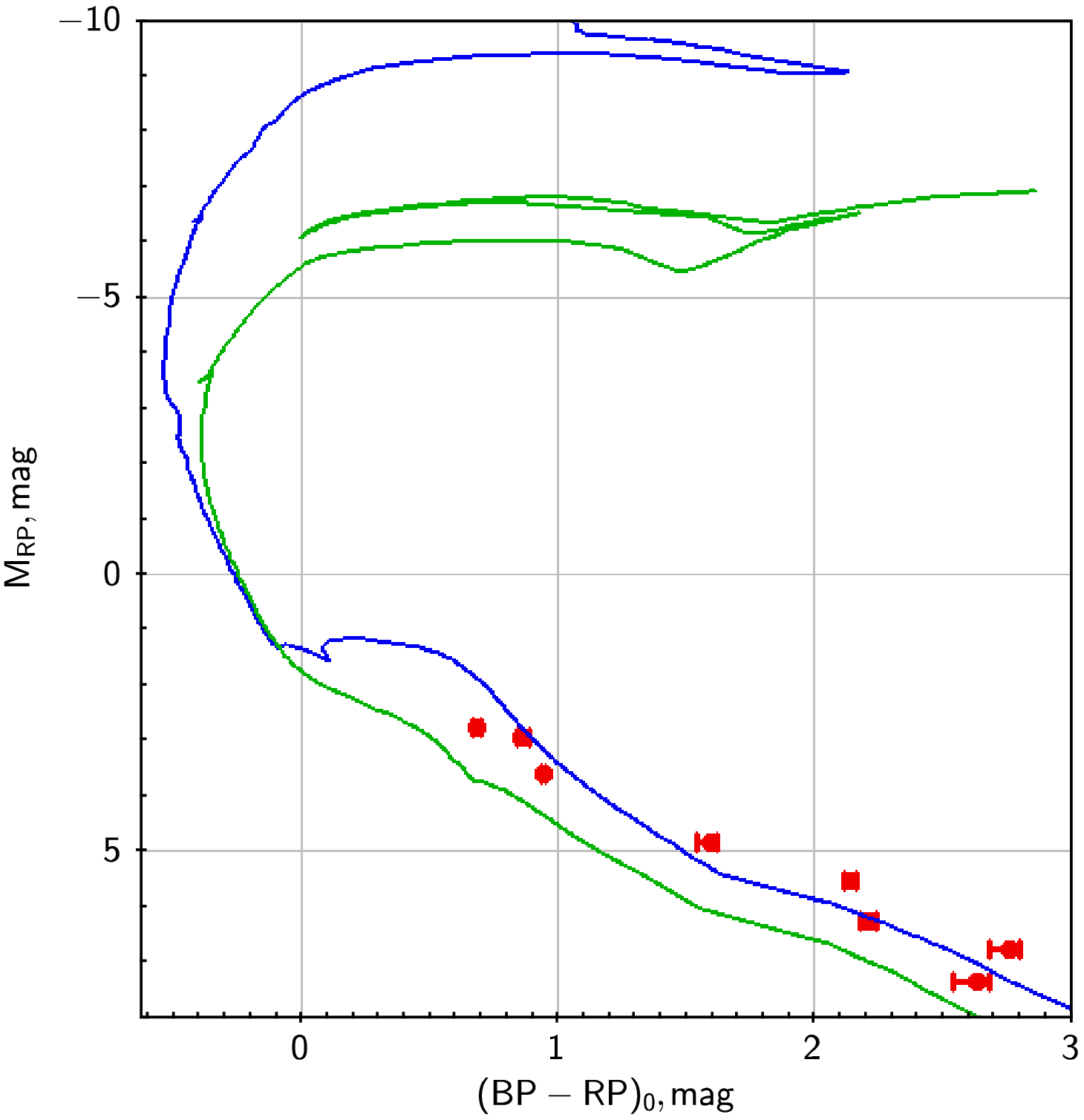} }}
    \caption{H-R diagrams for stars in the region MWSC 0739 taking into account a single value for the color excess (a) and derived from individual estimates of reddening for 8 stars (b) from the catalog of Ref. 14. The isochrones corresponding to ages logt=6.8 (blue curve) and 7.4 (green curve) are shown.}%
    \label{fig8}
\end{figure}

Also a very important circumstance is the photometric variability of 9 of the 17 stars of this group, detected in the GAIA observations (see Table 2, column Var), which is another sign of their low age.

\section{Discussion and conclusions}

Our results indicate definitely that the small open cluster MWSC 0739 is, in fact, a small group of extremely young stars of low and medium masses, of which a large fraction actually form a chain. Besides the visible stars, this group contains several IR-sources, including the presumed sources of the Herbig-Haro flows discussed above. The age of such objects, which still have dense dust shells, should be still lower. This indicates without doubt that after ``burst'' of star formation in the isolated dark cloud Dobashi 5006, which took place several million years ago, the process of forming new stars is still ongoing.

The reason for the onset of active star formation in this essentially isolated cloud is unclear. Perhaps in connection with this question it makes sense to point out that of the 8 stars from the group MWSC 0739 for which the DR3 Gaia catalog contains estimates of astrophysical parameters, two have levels of metallicity below the solar value, and three - much lower ([Fe/H] $<-0.7, < -1$, and even $-2$). A such difference seems strange for stars which in terms of spatial position and kinematics can be considered to be members of an open cluster. It may be explained by the inapplicability of the method used for determining the metallicity in GAIA to young objects with dust shells. To solve this problem, spectral observations of stars are highly desirable as they additionally make it possible to determine their radial velocities.

The main results of the present study can, therefore, be summarized as follows:

1.      In the region of the dark cloud Dobashi 5006, two new Herbig-Haro flows associated with IR-sources presumably belonging to Class I objects have been discovered

2.      The basic stellar-astronomical parameters (distance, proper motion, interstellar absorption) of objects in the cloud of the cluster MWSC 0739 have been estimated anew.

3.      This cluster is actually a small region of active star formation consisting of a group of at least 17 visible PMS-stars and several IR-sources.

\

This work was carried out as part of the thematic grant 21T-lC031 of the State committee for science of the Republic of Armenia. In this study we have used data obtained in the course of the Gaia survey of the European Space Agency processed by the DPAC consortium financed by the institutions participating in the multilateral Gaia collaboration.

In this work we have made active use of the SIMBAD and VIZIER data bases, and the Aladin virtual observatory developed by the Strasbourg Center for Stellar Data. The AKARI atlas is a project of the Japanese Agency for Aerospace studies, with participation by the European Space Agency. The PanSTARRS project is the result of a collaboration of the Hawaii Institute of Astronomy, the Lincoln Laboratory of M.I.T., the Maui High Performance Computing Center, and the International Corporation for Applied Science.

\

\
\textbf{References\\  }


1.      T. A. Movsessian, T. Yu. Magakian, and S. V. Dodonov, Mon. Not. Roy. Astron. Soc. 500, 2440 (2021).

2.      T. A. Movsessian, T. Yu. Magakian, and A. R. Andreasyan, Astrophysics 65, 193, (2022).

3.      K. Dobashi, Publ. Astron. Soc. Japan 63, SI (2011).

4.      N. V. Kharchenko, A. E. Piskunov, E. Schilbach, et al., Astron. Astrophys. 558, A53 (2013).

5.      A. L. Gyulbudagyan, R. Schwartz, and F. S. Nazaretyan, Soobshch. BAO 63, 3 (1990).

6.      S. N. Dodonov, S. S. Kotov, T. A. Movsesyan, et al., Astrophys. Bull. 72, 473 (2017).

7.      S. van den Bergh, Publ. Astron. Soc. Pacif. 87, 405 (1975).

8.      E. A. Magnier, A. W. Volp, K. Laan, et al., Astron. Astrophys. 352, 228 (1999).

9.      D. Froebrich, A. Scholz, and C. L. Raftery, Mon. Not. Roy. Astron. Soc. 374, 399 (2007).

10.     N. V. Kharchenko, A. E. Piskunov, E. Schilbach, et al., Astron. Astrophys. 585, A101 (2013).

11.     T. G. Sitnik, A. S. Rastorguev, A. A. Tatarnikova, et al., Mon. Not. Roy. Astron. Soc. 498, 5437 (2020).

12.     C. A. L. Bailer-Jones, J. Rybizki, M. Fouesneau, et al., Astron. J. 161, 147 (2021).

13.     F. Anders, A. Khalatyan, A. B. A. Queiroz, et al., Astron. Astrophys.658, A91 (2022).

14.     S. Wang and X. Chen, Astrophys. J.877 , id. 116 (2019).

\

\end{document}